\documentclass[11pt]{article}
\usepackage{moriond,epsfig}
\usepackage{amssymb}

\def\Journal#1#2#3#4{{#1} {\bf #2}, #3 (#4)}

\def\PRL{\em Phys. Rev. Lett.}
\def\PRD{{\em Phys. Rev.} D}

\def\ra{\rightarrow}

\def\be{\begin{equation}}
\def\ee{\end{equation}}
\def\bea{\begin{eqnarray}}
\def\eea{\end{eqnarray}}

\begin{document}
\rightline{TTP12-011}
\vspace*{4cm}
\title{GAMMA-RAY LINES CONSTRAINTS IN THE NMSSM\footnote{Talk at
\textit{Rencontres de Moriond, EW Interactions and Unified Theories}, Mar
4th-10th, 2012, La Thuile, Italy.}}

\author{Guillaume CHALONS}

\address{Institut f\"ur Theoretische Teilchenphysik, \\
Karlsruhe Institute of Technology, Universit\"at Karlsruhe \\
Engesserstra\ss{}e 7, 76128 Karlsruhe, Germany}

\maketitle\abstracts{
We present the computation of the loop-induced self-annihilation of dark matter
particles into two photons in the framework of the NMSSM. This process is a
theoretically clean observable with a ``smoking-gun'' signature but is
experimentally very challenging to detect. The rates were computed with the help
of the \texttt{SloopS} program, an automatic code initially designed for the
evaluation of processes at the one-loop level in the MSSM. We focused on a
light neutralino scenario and discuss how the signal can be enhanced in the
NMSSM with respect to the MSSM and then compared with the present limits given
by the dedicated search of the \texttt{FERMI-LAT} satellite on the monochromatic
gamma lines.}

\section{Introduction}
The existence of Cold Dark Matter (CDM) is the most compelling paradigm to
account for the wealth of cosmological and astrophysical data. Yet its
presence needs to be confirmed by its detection through direct or indirect
methods. With the advent of the LHC it has also started in collider
experiments. A theoretically well motivated and clean observable for the
indirect detection of dark matter are the so-called gamma-ray lines. Unlike
charged messengers, the photons are less affected by
astrophysical uncertainties and their propagation in the galaxy is easier to
model. In fact, pinning down the uncertainties on this detection channel boils
down to the modelisation of the galactic dark matter halo. Moreover no
astrophysical source is known to mimic this spectral feature, making this
signal as a ``smoking gun'' signature for the existence of Dark Matter (DM). On
the particle physics side, the prediction for the monochromatic gamma-ray lines
rates relies on the accurate computation of its self-annihilation cross section.
For the two gamma mode this cross section is generically very suppressed since
the CDM particle must be uncharged. Besides providing such a DM
candidate, Supersymmetry (SUSY) offers a fully computable framework. In one of
its extension, the NMSSM (\textit{Next-to-Minimal Supersymmetric Standard
Model}), this process is a loop-induced one (as in the most popular SUSY
incarnation, the MSSM) and requires the calculation of numerous Feynman diagrams
as well as an accurate calculation of loop integrals. The \texttt{SloopS} code
\cite{baro07,baro08,baro09} is such a tool which was designed initially for the
MSSM. Concerning DM studies it has been applied to the evaluation of the two
gamma mode in the MSSM \cite{boudjema05}, the NMSSM \cite{chalons11} and the
prediction of the relic density at Next-to Leading order (NLO)
\cite{baro07,chalons11,barosusy09} in the MSSM. In the present work we focus
on a light neutralino NMSSM DM candidate since a DM explanation is to be
advocated if one wants to explain recent direct detection measurements
\cite{cdmsII10,cogent10,cogent11,cresst11}. In the next section we will quickly
recap the NMSSM and highlight in which cases its DM phenomenology is peculiar
with respect to the MSSM together with an emphasis on the gamma-ray line
observable.

\section{Gamma-ray lines in the NMSSM}
\begin{figure}[b]
\begin{center}
\epsfig{file=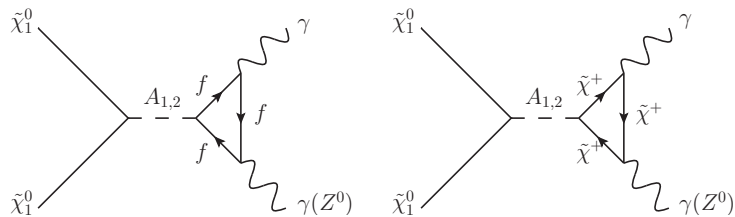,clip=,width=0.6\textwidth}
\caption{\label{NMSSMdiags} \em Additional NMSSM diagrams with s-channel
pseudoscalar exchange. The label $f$ stands for a SM fermion and $\chi^+$ to a
chargino.} 
\end{center}
\end{figure}

In the NMSSM the Higgs term of the superpotential involving the two Higgs
doublet is modified and a singlet term is added,
\begin{equation}
 W_{NMSSM} = W_{MSSM}^{\mu=0} + \lambda \hat S \hat H_u \hat H_d +
\frac{\kappa}{3}\hat S^3
\end{equation}
The VEV of the singlet generates an effective $\mu$ parameter with respect to
the MSSM, which is then naturally of order the EW
scale\cite{nmssmrev-ellwanger}. The soft-SUSY breaking Lagrangian is also
modified according to
\begin{equation}
 -\mathcal{L}_{\mathrm{soft}} = m^2_{H_u} |H_u|^2 +  m^2_{H_d} |H_d|^2 + m_S^2
|S|^2 +\left(\lambda A_\lambda H_u \cdot H_d S + \frac{1}{3} \kappa A_{\kappa}
S^3 + h.c\right)
\end{equation}
This has important phenomenological consequences, as compared to the MSSM, since
the Higgs and neutralino sector are extended due to the additional singlet
field and its fermionic superpartner, the singlino. Therefore the DM
phenomenology of the NMSSM has an extended parameter space region, the one
where the neutralino is mostly singlino, which makes it peculiar from the MSSM.
It has also been shown that it is easier to accommodate a light neutralino
$\tilde\chi_1^0$ and fulfill various kind of experimental constraints in the
NMSSM than the MSSM \cite{gunion-dmlight}. This is due to the presence of a more
``natural'' light pseudoscalar thanks to an approximate Peccei-Quinn symmetry.
The neutralinos can then efficiently annihilate through them and be a valid DM
candidate. A DM particle in the low mass region favours the indirect detection
channels (i.e the self-annihilation $\tilde\chi_1^0 \tilde\chi_1^0 \rightarrow
SM~SM$ in the galactic halo) since the signals are roughly inversely
proportional to $m_{\tilde\chi_1^0}$ and can be further enhanced in the NMSSM
through a light pseudoscalar $A_1$ s-channel resonance \cite{profumogg}. This
kind of mechanism can be at play for the $\gamma\gamma$ line signal and typical
NMSSM contributions for this process are shown in Fig.~\ref{NMSSMdiags}.
We can see that this process suffers from a loop suppression and therefore
generically the predicted rates are low. Hence, despite the fact that it is a
theoretically clean observable, it is experimentally very challenging to detect
these lines. It requires a very good rejection from the continuous gamma ray
background and also a fine energy resolution. The \texttt{FERMI-LAT} satellite
has a
dedicated search for these gamma-ray lines and up to now do not report any
observation, but limits on $\langle \sigma v \rangle_{\chi\chi \rightarrow
\gamma\gamma}$ were published instead \cite{fermigg}. . 
In the next section weinvestigate how these limits can constrain the light
pseudoscalar $A_1$ resonance in the low mass region ($m_\chi \lesssim 15$ GeV).
Obviously only the $\gamma\gamma$ channel is relevant here. The computation of
the $\gamma Z^0$ final state has been reported elsewhere \cite{chalons11}.

\section{\texttt{FERMI-LAT} constraints on the NMSSM parameter space}
\begin{figure}[htbp]
\begin{center}
 \epsfig{file=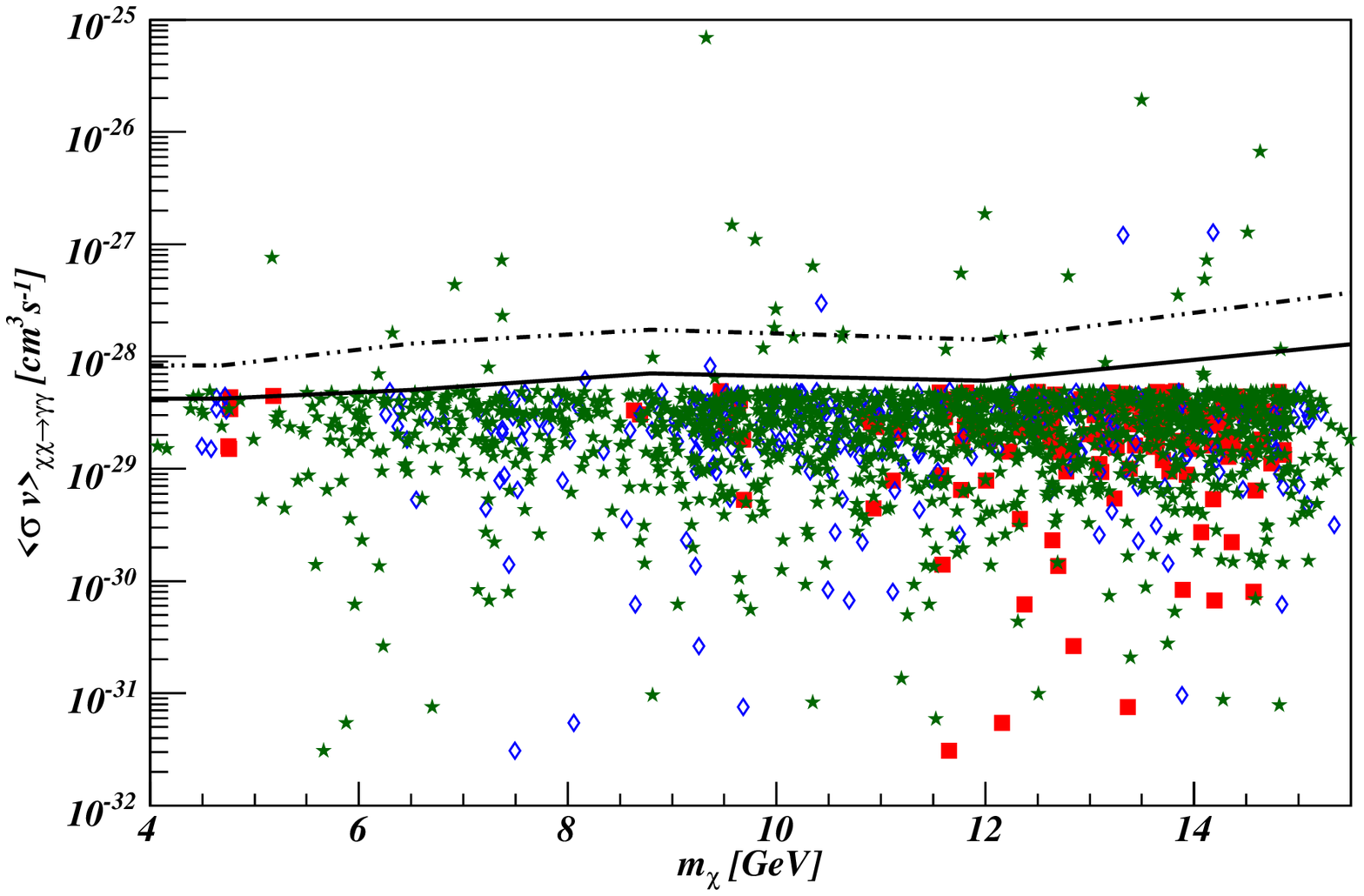,clip=,width=0.49\textwidth}
 \epsfig{file=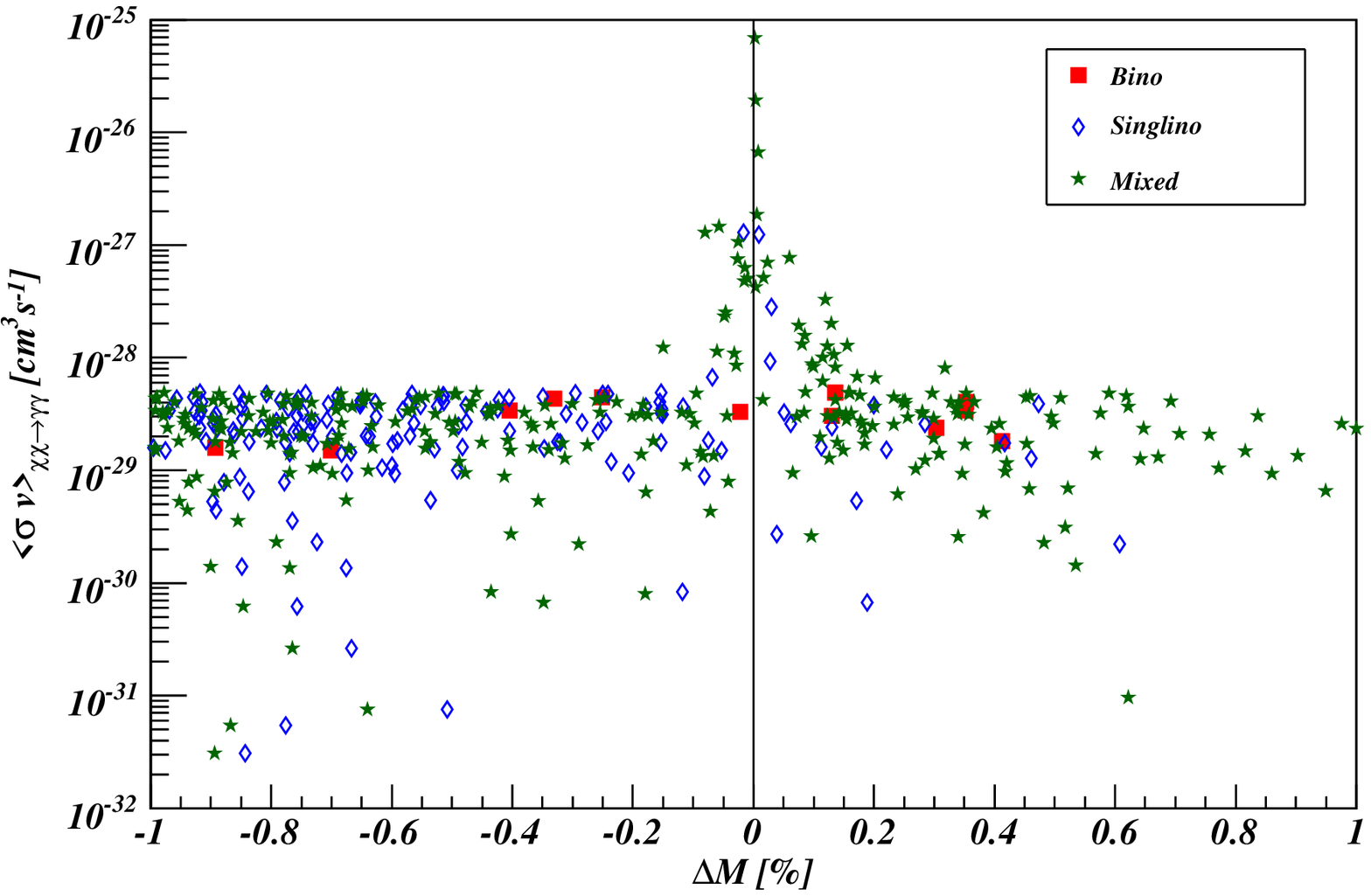,clip=,width=0.49\textwidth}
\caption{\label{scan} \em  $\langle \sigma v \rangle_{\chi\chi \rightarrow
\gamma\gamma}$with respect to the neutralino mass $m_\chi$ (left panel) and
$\Delta M$ (right panel, see text for its definition) for several types of
neutralinos. Solid lines are limit from the galactic center and the broken
one from the halo.}
\end{center}
\end{figure}

We performed a scan over the NMSSM parameter space focusing on a low mass
neutralino, as favoured by recent direct detection results
\cite{cdmsII10,cogent10,cogent11,cresst11} and computed the rate $\langle
\sigma v \rangle_{\chi\chi \rightarrow \gamma\gamma}$. However the published
\texttt{FERMI-LAT} limits do not extend to dark matter particles lighter than 30
GeV.
Nevertheless an analysis \cite{vertongen11} extended these limits down to 1
GeV, using \texttt{FERMI-LAT} data. These are the limits we used throughout this
work. The result of this scan is displayed in Fig.~\ref{scan}.
We can see on the left panel that current limits only exclude a minute portion
of the parameter space and the right panel shows that the excluded points
corresponds to situations where the annihilation occurs close to the resonance
by less than $\Delta M \leq |0.2|\%$, with $\Delta M = (2m_{\tilde\chi_1^0} -
m_{A_1})/m_{A_1}$. We can also observe that the highest rates are obtained when
the neutralino is significantly mixed. In details it is a mixture of
singlino-higgsino components, which is to be expected since these
enter the couplings of the neutralino to Higgses. We then investigate if the
limits on the spectral lines can further constrain the NMSSM parameter space. It
has been shown\cite{daniel-astrolim} that the 95\% limits of the
\texttt{FERMI-LAT} collaboration on the secondary gamma rays produced in dwarf
spheroidal galaxies (dSph) from the pair annihilation of DM particles into
quarks and/or taus can constrain the NMSSM parameter space. The authors of this
work provided us with 14 points of their MCMC scan giving a large pair
annihilation cross section but safe with respect to dSph limits and direct
detection searches. The points are sampled in each bins of $m_{\tilde\chi_1^0}$
between 1 and 15 GeV. We then used these input parameters to evaluate the rate
of the gamma-lines and if we could further constrain these scenarios. The
results are presented in Tab.~\ref{comparison}.
\begin{table}[htbp]
\caption{Comparison between dark matter annihilation into quarks
and/or taus and the loop-induced one into photons for each of the
bins between 1 and 14 GeV.\label{comparison}}
\begin{center}
\begin{tabular}{|c|c|c|c|c|c|c|c|}
\hline
$m_{\tilde\chi_1^0} [\mbox{GeV}]$ & 0.976 & 2.409 & 3.342 & 4.885 & 5.626 &
6.551 & 7.101\\
\hline
$\langle \sigma v \rangle_{\chi\bar\chi\ra q\bar q, \tau\bar\tau}
\times 10^{27}[\mbox{cm}^3\, \mbox{s}^{-1}]$ &0.209 &0.297 & 0.345& 3.298&
5.389&3.547 &2.425 \\
\hline 
$\langle \sigma v \rangle_{\chi\bar\chi\ra\gamma\gamma}\times
10^{32}[\mbox{cm}^3\, \mbox{s}^{-1}]$ & 0.008&0.267 &0.345 &0.262
&0.410& 0.427& 0.664\\
\hline
\end{tabular}

\begin{tabular}{|c|c|c|c|c|c|c|c|}
\hline
$m_{\tilde\chi_1^0}[\mbox{GeV}] $ & 8.513 & 9.274 & 10.27 & 11.50 & 12.74 &
13.51 & 14.48 \\
\hline$\langle \sigma v \rangle_{\chi\bar\chi\ra q\bar q, \tau\bar\tau}
\times 10^{27}[\mbox{cm}^3\, \mbox{s}^{-1}]$ & 2.161& 2.497& 2.323& 2.575&
3.224& 9.571&148.4 \\
\hline 
$\langle \sigma v \rangle_{\chi\bar\chi\ra\gamma\gamma}\times
10^{32}[\mbox{cm}^3\,
\mbox{s}^{-1}]$ & 0.220&0.665 &1.881 &2.456 &2.003 &17.49 &287.5 \\
\hline
\end{tabular}
\end{center}
\end{table}
The bottom line is that the predictions on the monochromatic $\gamma$ lines
rates are well below the present limits of \texttt{FERMI-LAT} and several orders
of magnitude of improvement on the experimental sensitivity are needed on this
observable to have a constraining power on these best fit points.

\section{Conclusion}
The mono-energetic gamma-ray line signal has spectacular features: a clear
”smoking-gun“ signature and points directly to the mass of the dark matter
particle. Moreover it do not suffer from astrophysical uncertainties and
depends only on the assumption of the dark matter halo. However discriminating
it from the overwhelming astrophysical background (supernov\ae{}, pulsars,
cosmic-rays...) remains an experimentally extremely challenging task. The
current sensitivity on the $\gamma\gamma$ mode permits to exclude only very
fine-tuned points where the LSP mass is close to the light pseudoscalar
resonance, when focusing on the low mass region. However the \texttt{FERMI-LAT}
mission is a long-termed one and the sensitivity is expected to be improved,
raising the possibility of excluding more featureless region of the NMSSM
parameter space. Finally a very recent independent paper\cite{weniger12} claimed
an indication for a gamma-ray line at 4.6$\sigma$ confidence level. We therefore
look forward a refined analysis of the \texttt{FERMI-LAT} for a confirmation of
this observation.

\section*{Acknowledgments}
G.C would like to thank the Moriond 2012 organisers for the possibility of
presenting this work. This work is supported by BMBF grant 05H09VKF.


\section*{References}
\begin{thebibliography}{99}
\bibitem{baro07}
N.~Baro, F.~Boudjema, A.~Semenov, \textit{Phys. Lett.} {\bf B660} (2008) 550,
  arXiv:0710.1821 [hep-ph].

\bibitem{baro08}
N.~Baro, F.~Boudjema, A.~Semenov, \Journal{\PRD}{78}{115003}{2008},
arXiv:0807.4668 [hep-ph].

\bibitem{baro09}
N.~Baro, F.~Boudjema, \Journal{\PRD}{80}{076010}{2009}, arXiv:0906.1665[hep-ph].

\bibitem{boudjema05}
F.~Boudjema, A.~Semenov, D.~Temes, \Journal{\PRD}{72}{055024}{2005},
hep-ph/0507127.

\bibitem{chalons11}
G.~Chalons, A.~Semenov, {\it JHEP} {\bf 1112}, 055 (2011) , arXiv:1110.2064
[hep-ph]

\bibitem{barosusy09}
N.~Baro, G.~Chalons, Sun~Hao, Proceedings of SUSY09, arXiv:0909.3263 [hep-ph].

\bibitem{chalons09}
N.~Baro, F.~Boudjema, G.~Chalons, Sun Hao, \Journal{\PRD}{81}{015005}{2010},
arXiv:0910.3293 [hep-ph].

\bibitem{cdmsII10}
[\texttt{CDMS-II} Collaboration], Z.~Ahmed {\it et al.}, {\it Science} {\bf
327}, 1619 (2010), arXiv:0912.3592 [astro-ph].

\bibitem{cogent10}
[\texttt{CoGeNT} Collaboration], C.~E.~Aalseth {\it et al.}, \Journal{\PRL}{106}
{131301}{2011}, arXiv:1002.4703. [astro-ph.CO].

\bibitem{cogent11}
[\texttt{CoGeNT} Collaboration], C.~E.~Aalseth {\it et al.},
\Journal{\PRL}{107}{141301}{2011}, arXiv:1106.0650 [astro-ph.CO].

\bibitem{cresst11}
[\texttt{CRESST} Collaboration], G.~Angloher {\it et al.}, arXiv:1109.0702
  [astro-ph.CO].

\bibitem{nmssmrev-ellwanger}
U.~Ellwanger, C.~Hugonie, A.~M.~Teixeira, {\it Phys. Rept.} {\bf 496}, 1
(2010), arXiv:0910.1785 [hep-ph].

\bibitem{gunion-dmlight}
J.~F.~Gunion, D.~Hooper, B.~McElrath, \Journal{\PRD}{73}{015011}{2006}.

\bibitem{profumogg}
F.~Ferrer, L.~Krauss, S.~Profumo, \Journal{\PRD}{74}{115007}{2006},
 hep-ph/0609257.

\bibitem{fermigg}
A. Abdo et al., \Journal{\PRL}{104}{091302}{2010},
\Journal{\PRL}{104}{091302}{2010}, arXiv:1001.4836
  [astro-ph.HE].

\bibitem{vertongen11}
G.~Vertongen, C.~Weniger, {\it JCAP} {\bf 1105}, 027 (2011), arXiv:1101.2610
  [hep-ph].

\bibitem{daniel-nmssm}
D.~A.~Vazquez, G.~Belanger, C.~Boehm, A.~Pukhov and J.~Silk,
\Journal{\PRD}{82}{115027}{2010}, arXiv:1009.4380 [hep-ph].

\bibitem{daniel-astrolim}
D.~A.~Vazquez, G.~Belanger, C.~Boehm, \Journal{\PRD}{84}{095008}{2011},
arXiv:1107.1614 [hep-ph].

\bibitem{weniger12}
C.~Weniger, arXiv:1204.2797 [hep-ph.]

\end{thebibliography}
\end{document}